\documentclass[12pt]{article}
\usepackage{fullpage}
\usepackage{cite}
\usepackage{amsmath}
\usepackage{amssymb}
\title{}
\date{}
\def\Pf { {\rm Pf} }
\def\tep { \tilde{\epsilon} }
\def\cAsc{ {\cal A}_{{\rm scalar}} }
\def\cAga{ {\cal A}_{\rm gauge} }
\def\cAgv{ {\cal A}_{\rm grav} }
\def\cA{ {\cal A} }
\def\para{\\ [-2mm]}
\def\barpsi{{\bar\psi}}
\def\be{\begin{equation}}
\def\ee{\end{equation}}
\def\ba{\begin{eqnarray}}
\def\ea{\end{eqnarray}}
\def\nl{\nonumber\\}
\def\eqn#1{eq.~(\ref{#1})} \def\Eqn#1{Equation~(\ref{#1})}
\def\eqns#1#2{eqs.~(\ref{#1}) and~(\ref{#2})}

\def\half{  {1\over 2} }

\def\SL2C{\mathrm{SL}(2,\mathbb{C})}

\def\id{  {\mathsf{1}\kern -3pt \mathsf{l} } }
\def\range{ 1,2, \cdots, n }
\def\psirange{ 1_\psi, 2, \cdots, n-1, n_\barpsi }
\def\psiphirange{ 1_\psi, 2_\phi, \cdots, n-1_\phi, n_\barpsi }

\def \Tr {\mathop{\rm Tr}\nolimits}

\begin{document}

\titlepage
\begin{flushright}
BOW-PH-159\\
\end{flushright}

\vspace{3mm}

\begin{center}
{\Large\bf\sf
Scattering equations 
and BCJ relations \\
for gauge and gravitational amplitudes \\ 
with massive scalar particles \\
}

\vskip 1.5cm

{\sc
Stephen G. Naculich\footnote{
Research supported in part by the National Science
Foundation under Grant No.~PHY10-67961.}
}

\vskip 0.5cm
{\it
Department of Physics\\
Bowdoin College\\
Brunswick, ME 04011, USA
}

\vspace{5mm}
{\tt
naculich@bowdoin.edu
}
\end{center}

\vskip 1.5cm

\begin{abstract}
We generalize the scattering equations to include
both massless and massive particles.
We construct an expression for the tree-level $n$-point amplitude
with $n-2$ gluons or gravitons and a pair of massive scalars
in arbitrary spacetime dimension
as a sum over the $(n-3)!$ solutions of the scattering equations,
\`a la Cachazo, He, and Yuan.
We derive the BCJ relations obeyed by these massive amplitudes.
\end{abstract}

\vspace*{0.5cm}

\vfil\break
\section{Introduction}
\setcounter{equation}{0}

Enormous interest has been generated over the last several years 
by the discovery of color-kinematic duality in gauge-theory amplitudes,
in particular because it allows for the construction of gravitational amplitudes
through the double-copy procedure \cite{Bern:2008qj,Bern:2010ue,Carrasco:2011hw}.
In their initial work,
Bern, Carrasco, and Johansson showed that the assumption of color-kinematic 
duality implies a previously unknown\footnote{The 
existence of these relations was presaged in the early 1980's in certain 
four-point amplitudes \cite{Zhu:1980sz,Goebel:1980es}.}
set of relations among 
tree-level color-ordered $n$-gluon amplitudes \cite{Bern:2008qj}.
These BCJ relations were subsequently proven using string-theory and field-theory 
techniques \cite{BjerrumBohr:2009rd,Stieberger:2009hq,Feng:2010my,Chen:2011jxa,Du:2011js}.
Although color-kinematic duality has not yet been proved at loop level, 
impressive evidence has been amassed to support the conjecture 
(see, e.g., 
refs.~\cite{Bern:2010ue,Carrasco:2011hw,Carrasco:2011mn,Bern:2012uf,Boels:2012ew,Bern:2013yya}).
Recent work of Cachazo, He, and Yuan (CHY)
has opened a new window on color-kinematic duality 
and the double-copy procedure by providing an alternative formula for 
tree-level gauge-theory and gravitational amplitudes 
in arbitrary spacetime dimension in terms of solutions of the scattering 
equations \cite{Cachazo:2013iaa,Cachazo:2013gna,Cachazo:2013hca,Cachazo:2013iea}.
This work has generated much interest \cite{Litsey:2013jfa,
Adamo:2013tca, Monteiro:2013rya, Mason:2013sva, Chiodaroli:2013upa,
Berkovits:2013xba, Dolan:2013isa, Adamo:2013tsa, Gomez:2013wza, Kalousios:2013eca,
Stieberger:2014hba, Yuan:2014gva, Weinzierl:2014vwa, Dolan:2014ega, 
Bjerrum-Bohr:2014qwa,He:2014wua, Kol:2014yua, Kol:2014zca, Geyer:2014fka,
Naculich:2014rta}
and has been utilized in proofs \cite{Schwab:2014xua,
Afkhami-Jeddi:2014fia,Zlotnikov:2014sva,Kalousios:2014uva}
of a new soft graviton theorem \cite{Cachazo:2014fwa}.
\para

Previous studies have focused on scattering amplitudes
for massless particles transforming in the adjoint representation
of the gauge group.
In this paper, we explore color-kinematic duality 
of gauge-theory amplitudes involving massive particles 
and in representations other than the adjoint.\footnote{Recent 
work on color-kinematic duality for other representations includes 
refs.~\cite{Vera:2012ds,Johansson:2013nsa,Chiodaroli:2013upa,Johansson:2014zca}.} 
Specifically, we examine the class of amplitudes 
consisting of $(n-2)$ gluons and 
a pair of massive particles $\psi$ of arbitrary spin
transforming in the fundamental representation of the gauge group.
These amplitudes can be decomposed into $(n-2)!$ color-ordered amplitudes
\ba
\cA (\psirange)
&=& 
\sum_{\gamma \in S_{n-2} }
{\bf t}_{1 \gamma n} ~A(1_\psi, \gamma(2),  \cdots, \gamma(n-1), n_\barpsi )  \,,
\label{introAtA}
\\
{\bf t}_{1 \gamma n}
&=&
\left( {T}^{\textsf{a}_{\gamma(2)}}{T}^{\textsf{a}_{\gamma(3)}}
\cdots {T}^{\textsf{a}_{\gamma(n-1)}} \right)^{\textsf{i}_1}_{~~ \textsf{i}_n} 
\label{deft}
\ea
where $T^\textsf{a}$ denote generators in the fundamental representation.
\para

The BCJ relations that apply to tree-level color-ordered $n$-gluon amplitudes
hinge on two key ingredients:
color-kinematic duality, and 
the properties of the propagator matrix \cite{Vaman:2010ez}.
The rank of the propagator matrix fixes the number of BCJ relations,
and the form of these relations is determined by the null eigenvectors of this matrix.
In sec.~\ref{sec:propagator} of this paper, 
we show that the propagator matrix for the amplitude (\ref{introAtA})
has the same rank as that for the $n$-gluon amplitude,
and that its null eigenvectors imply that the color-ordered amplitudes obey 
\be
0 = \sum_{a=3}^n \left(- m_\psi^2 +\sum_{b=a}^n s_{2b} \right) 
A(1_\psi, 3, \cdots, a-1, 2, a, \cdots, n_\barpsi) 
\label{intromassivebcj}
\ee
where $m_\psi$ is the mass of $\psi$,
provided that the amplitude satisfies color-kinematic duality.
We check  \eqn{intromassivebcj} using various results in the literature,
providing evidence for the assumption of color-kinematic duality for the
class of amplitudes we are considering.
\para

In this paper, we also propose a generalization of the scattering 
equations \cite{Cachazo:2013iaa,Cachazo:2013gna,Cachazo:2013hca,Cachazo:2013iea}
to massive particles
\be
\sum_{b\neq a} \frac{k_a \cdot k_b + \Delta_{ab} } {\sigma_{a}-\sigma_{b}} = 0 , 
\qquad \sigma_a \in \mathbb{C} \mathbb{P}^1,
\qquad    a = 1, \cdots, n
\label{introscatt}
\ee
where 
\be
\Delta_{ab} = \Delta_{ba}, \qquad \sum_{b\neq a} \Delta_{ab} = m_a^2 
\label{introconstraints}
\ee
is imposed to guarantee $\SL2C$ invariance of the equations.
We then use these scattering equations to 
construct amplitudes in a trio of theories:

\begin{itemize}

\item
a double-color theory of massless and massive scalars belonging 
to the adjoint and fundamental representations of $U(N) \times U(\tilde{N})$,

\item 
a gauge theory of gluons and massive scalars in the fundamental representation,
and

\item a gravitational theory of gravitons and massive scalars.
\end{itemize}

We present a CHY-type formula 
for the amplitudes of $(n-2)$ massless and two massive scalars 
of the double-color theory
in terms of a sum over solutions of the massive scattering equations
(\ref{introscatt}). 
We establish its validity by proving that 
the double-partial amplitudes 
agree with the propagator matrix for the amplitudes (\ref{introAtA})
computed in sec.~\ref{sec:propagator}.
\para

We present a CHY-type formula 
for the amplitudes of $(n-2)$ gluons and two massive scalars 
in the fundamental representation,
and check agreement with previously-known results.
We explicitly show that the color-ordered amplitudes
satisfy the BCJ relations (\ref{intromassivebcj}).
In general all solutions of the scattering equations contribute
to each amplitude.
In four dimensions and for massless scalars, however,
we observe that only a subset of solutions contributes to the amplitude,
the subset depending on the helicities of the gluons.
\para

We also propose a CHY-type formula
for amplitudes of $(n-2)$ gravitons and two massive scalars,
and verify agreement with the known four-point amplitude.
\para

This paper is structured as follows:
in sec.~\ref{sec:propagator}, 
we demonstrate the relationship between 
the propagator matrices for the $n$-gluon amplitude and 
for the amplitude for $(n-2)$ gluons and two massive fundamentals.
We use this to derive BCJ relations for the latter, 
assuming color-kinematic duality.
In sec.~\ref{sec:scatt}, 
we generalize the scattering equations to massive external particles.
In secs.~\ref{sec:doublecolor}, \ref{sec:gauge}, and  \ref{sec:gravity}, 
we construct the amplitudes in 
double-color, gauge, and gravity theories respectively.
Sec.~\ref{sec:concl} contains conclusions and directions for further work.

\section{BCJ relations for  massless and massive amplitudes} 
\setcounter{equation}{0}
\label{sec:propagator}

The number and form of the BCJ relations among color-ordered amplitudes 
are completely fixed by the rank and null eigenvectors of the propagator matrix.
In this section, we explicitly define this matrix,
and determine the relationship between the propagator matrices for 
the $n$-gluon amplitude and  the amplitude for 
$(n-2)$ gluons and a pair of massive fundamentals.
This allows us to derive the form of the BCJ relations 
among massive amplitudes, provided color-kinematic duality is satisfied.

\subsection{BCJ relations for $n$-gluon amplitudes}

We begin by reviewing the BCJ relations for the tree-level $n$-gluon 
amplitude in the spirit of ref.~\cite{Vaman:2010ez}. 
This amplitude can be expressed as a sum over the $(2n-5)!!$ diagrams  
that can be assembled from cubic vertices
\cite{Bern:2008qj}
\be
\cA  (\range)
~=~ \sum_i 
{c_i ~ n_i 
\over d_i } \,.
\label{Acnd}
\ee
We define the {\it backbone} of a diagram 
as the path from the first to the $n$th external line.
All of the other external lines attach to the backbone 
either directly or via side branches. 
The subset of $(n-2)!$ diagrams 
with no side branches
(i.e. all external lines 
emerge directly from the backbone)
we refer to as half-ladder diagrams.
\para

Associated with each diagram $i$ is a color factor $c_i$
obtained by sewing together three-gluon\footnote{The
four-gluon vertex is expressed in terms of a 
linear combination of products of three-gluon factors 
$ f^{\textsf{a} \textsf{b} \textsf{e}} f^{\textsf{c} \textsf{d} \textsf{e}},$
$f^{\textsf{a} \textsf{c} \textsf{e}} f^{\textsf{d} \textsf{b} \textsf{e}},$ and
$f^{\textsf{a} \textsf{d} \textsf{e}} f^{\textsf{b} \textsf{c} \textsf{e}}$,
which is why the contribution from Feynman diagrams containing quartic vertices 
can be parceled out among several purely cubic color factors.}
vertices $f^{\textsf{abc}}$.
Among these are the color factors ${\bf c}_{1 \gamma n }$
associated with half-ladder diagrams 
\be
{\bf c}_{1 \gamma n } ~\equiv~
{\bf c}_{1 \gamma(2) \cdots \gamma(n-1) n } 
~\equiv~  \sum_{\textsf{b}_1,\ldots,\textsf{b}_{n{-}3}} 
f^{\textsf{a}_{1} \textsf{a}_{\gamma(2)} \textsf{b}_1}
f^{\textsf{b}_{1} \textsf{a}_{\gamma(3)} \textsf{b}_2}\cdots 
f^{\textsf{b}_{n{-}3} \textsf{a}_{\gamma(n{-}1)} \textsf{a}_{n}} 
\ee
where $ \gamma $ denotes a permutation of $\{2, \cdots, n-1\}$. 
An arbitrary color factor  $c_i$
can be written as a linear combination
of half-ladder color factors 
by repeatedly applying the Jacobi identity 
$ f^{\textsf{a} \textsf{b} \textsf{e}} f^{\textsf{c} \textsf{d} \textsf{e}}
+f^{\textsf{a} \textsf{c} \textsf{e}} f^{\textsf{d} \textsf{b} \textsf{e}}
+f^{\textsf{a} \textsf{d} \textsf{e}} f^{\textsf{b} \textsf{c} \textsf{e}}=0 $
to the side branches,
starting from the backbone and working outward,
as described in ref.~\cite{DelDuca:1999rs}.
Thus, the set of $(n-2)!$ half-ladder color factors
forms an independent basis for the color factors\footnote{The 
coefficients $M_{i, \alpha}$ can be computed by using 
$f^{\textsf{abc}} = \Tr(T^\textsf{a} [ T^\textsf{b}, T^\textsf{c}] )$
to decompose $c_i$  into a linear combination of  traces
$\Tr[\alpha]\equiv
{\rm Tr}({T}^{\textsf{a}_{\alpha(1)}}{T}^{\textsf{a}_{\alpha(2)}}\cdots {T}^{\textsf{a}_{\alpha(n)}})$,
and reading off the coefficients;
see, e.g., ref.~\cite{Naculich:2014rta}.}
\be
c_i ~=~ \sum_{\gamma \in S_{n-2}}  M_{i, 1\gamma n} {\bf c}_{1 \gamma n } 
\label{cMc}
\ee
and can be used to decompose the $n$-gluon amplitude (\ref{Acnd}) as \cite{DelDuca:1999ha,DelDuca:1999rs}
\ba
\cA (\range)  
&=& 
\sum_{\gamma \in S_{n-2}}  
{\bf c}_{1 \gamma n } ~A(1, \gamma(2), \cdots, \gamma(n-1), n)  \,,
\label{AcA}
\\
A(1, \gamma(2),  \cdots, \gamma(n-1), n )  
&=&
\sum_i 
{M_{i, 1\gamma n}  ~ n_i 
\over d_i } 
\label{Amnd}
\ea
where $A(1, \gamma(2), \cdots, \gamma(n-1), n) $ are the color-ordered amplitudes 
belonging to the Kleiss-Kuijf basis.\footnote{All other color-ordered amplitudes are related to these by the 
Kleiss-Kuijf relations \cite{Kleiss:1988ne,DelDuca:1999rs}.}
\para

The kinematic numerators $n_i$ associated with each diagram $i$ 
are functions of the momenta and polarizations of the external gluons.
The hypothesis of color-kinematic duality is that 
the kinematic numerators $n_i$ obey the same Jacobi relations as $c_i$, 
and thus 
can similarly be expressed in terms of 
$(n-2)!$ half-ladder numerators ${\bf n}_{1 \gamma n }$:
 \be
n_i ~=~ \sum_{\gamma \in S_{n-2}}  M_{i, 1\gamma n} 
~{\bf n}_{1 \gamma n } \,.
\label{nMn}
\ee
Using \eqn{nMn}, the color-ordered amplitude (\ref{Amnd}) can be written
\be 
A(1, \gamma(2), \cdots, \gamma(n-1), n)  
~=~  \sum_{\delta  \,\in\, S_{n-2}} 
m( 1 \gamma n| 1 \delta n) ~{\bf n}_{1 \delta n }   
\label{Amn}
\ee
where we define the propagator matrix
\be 
m( 1 \gamma n| 1 \delta n) 
= \sum_i 
{M_{i,1\gamma n} M_{i, 1\delta n} \over d_i} 
\label{propagator}
\ee
as the sum (weighted by the denominators $1/d_i$ of the diagrams) 
over those cubic diagrams that contribute to both $\Tr[1 \gamma n]$ and $\Tr[1 \delta n]$.
\para

Although not obvious {\it a priori}, 
the $(n-2)! \times (n-2)!$ propagator matrix $ m( 1 \gamma n| 1 \delta n) $
has rank $(n-3)!$  as a consequence of momentum conservation \cite{Vaman:2010ez}. 
In the scattering equation approach, 
$ m( 1 \gamma n| 1 \delta n) $
can be interpreted as a {\it double-partial amplitude}
in a theory of scalar particles transforming in the adjoint 
representation of the group $U(N) \times U(\tilde{N})$ \cite{Cachazo:2013iea}.
The matrix of double-partial amplitudes can be expressed in terms 
of the $(n-3)!$ independent solutions of the scattering equations, 
which makes its reduced rank manifest.
\para

The reduced rank of the propagator matrix
implies that it possesses $(n-2)!-(n-3)!$ null eigenvectors.
Consequently, the $(n-2)!$ Kleiss-Kuijf color-ordered amplitudes  (\ref{Amn})
obey an equal number of independent relations.
All of these BCJ relations can be generated by the 
fundamental BCJ relation (and permutations thereof)
\cite{BjerrumBohr:2009rd,Feng:2010my,Sondergaard:2011iv}
\be
0 = \sum_{a=3}^n \left( \sum_{b=a}^n s_{2b}\right) A(1, 3, \cdots, a-1, 2, a, \cdots, n) 
\label{fundbcj}
\ee
where $s_{ab} \equiv (k_a + k_b)^2$
and $k_a$ are the momenta of the external particles.
\para

We close by emphasizing that the number and form of the BCJ relations 
are entirely determined by the propagator matrix, 
independent of the expressions for the kinematic numerators $n_i$
(provided only that the latter obey color-kinematic duality).

\subsection{BCJ relations for amplitudes with massive particles}

Next we turn to tree-level amplitudes for $(n-2)$ gluons 
and a pair of massive fundamentals $\psi$ of arbitrary spin. 
Again, we can express this as a sum over cubic diagrams
\be
\cA (\psirange)
~=~ \sum_i 
{c'_i ~ n'_i 
\over d'_i } 
\label{Acndprime}
\ee
where we decorate the color factors, kinematic numerators, and denominators with primes
to distinguish them from the analogous quantities for $n$-gluon amplitudes.
These diagrams are in one-to-one correspondence with the $n$-gluon diagrams,
in which the backbone of each $n$-gluon diagram is replaced by 
a string of propagators of massive fundamentals.
\para

The color factor $c'_i$ associated with each new diagram is 
obtained by sewing together cubic\footnote{As before, 
four-gluon vertices can be parceled out among 
three separate pairs of cubic vertices.
For spin-half $\psi$, there are no $\barpsi g g \psi $ vertices,
whereas for spin-zero $\psi$, 
the $\barpsi g g \psi $ vertices are proportional to 
$ \left\{  T^{\textsf{a}}, T^{\textsf{b}} \right \}^{\textsf{i}}_{~\textsf{j}} $
and so can be recast as a pair of 
$\barpsi g \psi$ vertices.}
$ggg$ vertices $f^{\textsf{abc}}$
and $\barpsi g \psi $ vertices $(T^{\textsf{a}})^{\textsf{i}}_{~\textsf{j}} $.
As in the case of $n$-gluon diagrams, 
each color factor $c_i'$ can be reduced to a linear combination of half-ladder color factors
${\bf t}_{1 \gamma n}$, defined in \eqn{deft}, 
by repeatedly applying\footnote{This 
was also used recently in ref.~\cite{Johansson:2014zca}.}
\be
f^{\textsf{abc}} \left( T^{\textsf{c}} \right)^{\textsf{i}}_{~\textsf{j}} 
= \left[ T^{\textsf{a}}, T^{\textsf{b}} \right]^{\textsf{i}}_{~\textsf{j}} 
\ee
to any gluon propagator emerging from the backbone,
until all the factors of $f^{\textsf{abc}}$ are removed from $c_i'$.
This process results in the decomposition
\be
c'_i ~=~ \sum_{\gamma \in S_{n-2}}  M_{i, 1\gamma n} {\bf t}_{1 \gamma n } 
\ee
where the coefficients 
$M_{i, 1\gamma n}$ are precisely the same as in the $n$-gluon case.
The ${\bf t}_{1 \gamma n } $ 
can thus be used to decompose the amplitude (\ref{Acndprime}) into color-ordered amplitudes 
\ba
\cA (\psirange)
&=&
\sum_{\gamma \in S_{n-2} }
{\bf t}_{1 \gamma n} ~A(1_\psi, \gamma(2),  \cdots, \gamma(n-1), n_\barpsi )  \,,
\label{AtA}
\\
A(1_\psi, \gamma(2),  \cdots, \gamma(n-1), n_\barpsi )  
&=&
\sum_i 
{M_{i, 1\gamma n}  ~ n'_i 
\over d'_i }  \,.
\label{Amndprime}
\ea
To discover whether the color-ordered amplitudes 
(\ref{Amndprime}) satisfy relations analogous to those for $n$-gluon amplitudes,
we must determine 
(a) whether color-kinematic duality continues to hold, and
(b) whether the propagator matrix continues to possess null eigenvectors 
as a consequence of momentum conservation 
even when some of the particles are massive.
\para

Four-point amplitudes of gluons and massive fundamental fields 
were examined in refs.~\cite{Zhu:1980sz,Goebel:1980es} 
and it was shown that the kinematic numerators $n_i'$ obey algebraic
relations analogous to $c_i'$ for both 
spin-zero and spin-half fundamentals.
We will therefore proceed to assume that this condition
is satisfied for higher-point amplitudes to examine the consequences.
In that case, we can write 
\be
n'_i ~=~ \sum_{\gamma \in S_{n-2}}  M_{i, 1\gamma n} {\bf n'}_{1 \gamma n } 
\ee
implying that
\be
A(1_\psi, \gamma(2), \cdots, \gamma(n-1), n_\barpsi)  
~=~  \sum_{\delta  \,\in\, S_{n-2}} 
m'( 1 \gamma n| 1 \delta n) ~{\bf n'}_{1 \delta n }  
\label{Amnprime}
\ee
where 
\be
m'( 1 \gamma n| 1 \delta n) 
= \sum_i 
{M_{i,1\gamma n} M_{i, 1\delta n} \over d'_i} \,.
\label{primepropagator}
\ee
The null eigenvectors of 
\eqn{primepropagator}
therefore determine the (BCJ) relations among the color-ordered 
amplitudes (\ref{Amnprime}),
provided that color-kinematic duality is satisfied.
\para

Observe that the propagator matrix
(\ref{primepropagator})
is the same as that for the $n$-gluon amplitude  (\ref{propagator})
except that the denominators  $d_i'$  must be adjusted 
to account for the mass of $\psi$.   
The denominator of each diagram consists of a product of inverse propagators.
Each inverse propagator belonging to a side branch
is of the form $(\sum_{a \subset S}  k_a)^2$, 
where $S$ is some subset of the gluon momenta
$\{ k_2, \cdots, k_{n-1} \}$.
Since $k_a^2=0$ for the gluons, 
this consists of a sum of terms $k_a \cdot k_b$
where $2 \le a, b \le n-1$,
which are the same for $d_i$ and $d'_i$.
Each inverse propagator belonging to the backbone of fundamental fields
is of the form 
\be
 \left(k_n + \sum_{a \subset S}  k_a\right)^2 - m_\psi^2
= \left(\sum_{a \subset S}  k_a\right)^2  + \sum_{a \subset S} 2 k_a \cdot k_n  \,.
\ee 
Thus, when $d_i'$ is expressed in terms of 
$k_a \cdot k_b$ with $2 \le a <  b \le n$
(eliminating $k_1$ if necessary
by using momentum conservation $\sum_{a=1}^n k_a=0$),
the dependence on $m_\psi$ disappears\footnote{The 
remaining constraint among this set of variables
$\sum_{2 \le a < b \le n} k_a \cdot k_b = 0$
is also independent of $m_\psi$.},
so $d_i'$ is identical to $d_i$.
Consequently, 
propagator matrix $ m'( 1 \gamma n| 1 \delta n)  $ 
when expressed in terms of
these same variables\footnote{Naturally,
one can alternatively express $d_i'$ in terms of
$k_a \cdot k_b$ with $1 \le a <  b \le n-1$,
eliminating $k_n$ using momentum conservation.}
is identical to the $n$-gluon propagator matrix
$ m( 1 \gamma n| 1 \delta n)$.
\para

As a result, we may obtain the BCJ relations for the
amplitudes of $(n-2)$ gluons and two fundamentals 
in terms of those for the $n$-gluon amplitude.
Expressed in terms of $k_a \cdot k_b$ with $2 \le a <  b \le n$, 
the fundamental BCJ relation (\ref{fundbcj}) for the $n$-gluon amplitude is
\be
0 =  \sum_{a=3}^n \left( \sum_{b=a}^n 2 k_2 \cdot k_b \right) 
A(1, 3, \cdots, a-1, 2, a, \cdots, n) \,.
\ee
Since the propagator matrix $ m'( 1 \gamma n| 1 \delta n)  $ 
in these variables has same form as $ m( 1 \gamma n| 1 \delta n)  $,
so do their null eigenvectors,
and therefore the fundamental BCJ relation for the massive amplitude is 
\be
0 =  \sum_{a=3}^n \left( \sum_{b=a}^n 2 k_2 \cdot k_b \right) 
A(1_\psi, 3, \cdots, a-1, 2, a, \cdots, n_\barpsi) \,.
\label{massivebcj}
\ee
When rewritten in terms of $s_{ab}$ this becomes 
\be
0 = \sum_{a=3}^n \left( -m_\psi^2 + \sum_{b=a}^n s_{2b}\right) 
A(1_\psi, 3, \cdots, a-1, 2, a, \cdots, n_\barpsi) \,.
\label{massivebcjfinal}
\ee
This is the fundamental BCJ relation obeyed  by tree-level amplitudes with 
$(n-2)$ gluons and two massive fundamentals $\psi$,
provided that color-kinematic duality is satisfied. 
\para

For $n=4$, 
\eqn{massivebcjfinal} becomes
\be 
(s_{12}-m_\psi^2) A(1_\psi,2,3,4_\barpsi) = (s_{13}-m_\psi^2) A(1_\psi,3,2,4_\barpsi)
\label{zhubcj}
\ee
which was established in refs.~\cite{Zhu:1980sz,Goebel:1980es} for both spin-zero 
and spin-half fundamentals.
We have also verified  \eqn{massivebcjfinal}
for various five- and six-point amplitudes with massive scalars,
using known results in four 
dimensions \cite{Bern:1996ja,Badger:2005zh,Forde:2005ue},
e.g. eqs.~(\ref{spinorfiveA}-\ref{spinorfiveC}) for $n=5$ 
and \eqn{spinorsix}  for $n=6$.
This provides evidence for the assumption of color-kinematic duality 
for this class of amplitudes.

\section{Scattering equations}
\setcounter{equation}{0}
\label{sec:scatt}

The equations\footnote{These equations have appeared previously 
in a string-theory context \cite{Fairlie:1972zz,Gross:1987kza,Fairlie:2008dg}.}
\be
\sum_{b\neq a} \frac{k_a \cdot k_b} {\sigma_{a}-\sigma_{b}} = 0 , 
\qquad \sigma_a \in \mathbb{C} \mathbb{P}^1,
\qquad    a = 1, \cdots, n
\label{scatt} 
\ee
dubbed {\it scattering equations} by Cachazo, He, and Yuan (CHY),
have recently proven to play a key role 
in the structure of scattering amplitudes of massless particles 
\cite{Cachazo:2013iaa,Cachazo:2013gna,Cachazo:2013hca,Cachazo:2013iea}.
The set of equations (\ref{scatt}) is invariant under 
$\SL2C$  transformations
\be
\sigma \longrightarrow {A \sigma + B \over C \sigma + D}, \qquad A D-B C = 1
\label{sl2c}
\ee
provided $\sum_{a=1}^n k_a =0$ and $k_a^2=0$.
Equivalently, only $n-3$ of the $n$ equations (\ref{scatt})
are independent.
In refs.~\cite{Cachazo:2013gna,Cachazo:2013hca},
CHY presented novel expressions for tree-level scattering amplitudes 
of gluons and gravitons in terms of sums over the solutions of \eqn{scatt}.
\para

In order to apply the CHY approach to the class of 
amplitudes considered in this paper,
we generalize the scattering equations 
to the case where the external particles are massive, $k_a^2 = m_a^2$.
(Dolan and Goddard have previously considered a generalization 
in which all external masses are equal \cite{Dolan:2013isa,Dolan:2014ega}.)
We propose modifying \eqn{scatt} to  
\be
f_a = 0, \qquad    a = 1, \cdots, n
\label{newscatt}
\ee
where 
\be
f_a  \equiv \sum_{b\neq a} \frac{k_a \cdot k_b + \Delta_{ab}} {\sigma_{a}-\sigma_{b}}, 
\qquad    a = 1, \cdots, n
\ee
with 
\be
\Delta_{ab} = \Delta_{ba}, \qquad \sum_{b\neq a} \Delta_{ab} = m_a^2 \,.
\label{constraints}
\ee
It is straightforward to verify that
$f_a \to (C \sigma_a+ D)^2 f_a$ under  \eqn{sl2c},
so that the modified scattering equations (\ref{newscatt})
remain invariant under $\SL2C$ transformations.
Furthermore, the three linear combinations 
\be
\sum_{a=1}^n f_a, \qquad 
\sum_{a=1}^n \sigma_a f_a, \qquad
\sum_{a=1}^n \sigma_a^2 f_a
\ee
vanish identically precisely when \eqn{constraints} is satisfied, 
which implies that only $n-3$ of the massive scattering equations (\ref{newscatt})
are independent.
\para

Dolan and Goddard showed that when all external particles have equal mass $m$, 
the scattering equations are generalized to the form (\ref{newscatt}) 
with $\Delta_{ab} = \half m^2  (\delta_{a+1,b} + \delta_{a-1,b})$,
which indeed satisfies \eqn{constraints}.
They observe, however, that this choice imposes a specific ordering on the $n$ particles, 
breaking the permutation invariance of the massless 
equations \cite{Dolan:2013isa,Dolan:2014ega}.
\para

For the case of the amplitudes considered in this paper, 
in which only two of the external particles are massive 
($m_1^2 = m_n^2 = m_\psi^2$ and $m_a=0$ for $a=2, \cdots, n-1$),
the constraints (\ref{constraints})
are rather naturally satisfied by choosing 
$\Delta_{1n}= \Delta_{n1} = m_\psi^2$ with all other
$\Delta_{ab}$ vanishing. 
Specifically, the massive scattering equations (\ref{newscatt}) become
\begin{align}
\sum_{b=2}^{n-1} \frac{k_1 \cdot k_b} {\sigma_{1}-\sigma_{b}}
+ \frac{k_1 \cdot k_n + m_\psi^2 } {\sigma_{1}-\sigma_{n}}
&=0, 
\\
\sum_{b\neq a} \frac{k_a \cdot k_b } {\sigma_{a}-\sigma_{b}} 
&=0 , \quad {\rm for} \quad a = 2, \cdots, n-1
\label{middle}
\\
\frac{k_n \cdot k_1 + m_\psi^2 } {\sigma_{n}-\sigma_{1}}
+ \sum_{b=2}^{n-1} \frac{k_n \cdot k_b  } {\sigma_{n}-\sigma_{b}}
&=0 . 
\end{align}
Despite the presence of $m_\psi^2$, 
these equations are essentially equivalent to the massless scattering equations.
This can be seen by using momentum conservation to eliminate $k_1$
(alternatively, $k_n$).
The equations,
when expressed in terms of the variables 
$k_a \cdot k_b$ with $2 \le a <  b \le n$
(alternatively, $1 \le a <  b \le n-1$),
are identical to the massless scattering equations, 
and therefore have the same set of solutions.\footnote{Again, 
we note the fact that the remaining constraint 
$\sum_{2 \le a < b \le n} k_a \cdot k_b = 0$
among this set of variables is independent of $m_\psi$.}
\para

We now use these equations to construct amplitudes for 
scalar, gauge, and gravitational theories with massive particles.

\section{Double-color amplitudes} 
\setcounter{equation}{0}
\label{sec:doublecolor}

In ref.~\cite{Cachazo:2013iea},
Cachazo, He, and Yuan presented a new formulation 
for the tree-level amplitudes of three interrelated theories
in terms of the solutions of the massless scattering 
equations (\ref{scatt}).
Their unified formula computes the
$n$-point amplitudes of colored scalars, of gluons,
and of gravitons.
The scalar theory is the simplest,
and contains massless scalar particles 
$ \phi^{\textsf{a}\textsf{a'}}$
in the adjoint of the color group $U(N) \times U(\tilde{N})$ 
with cubic interactions of the form
\be
f^{\textsf{abc}}\tilde f^{\textsf{a'b'c'}}
\phi_{\textsf{a}\textsf{a'}}\phi_{\textsf{b}\textsf{b'}}\phi_{\textsf{c}\textsf{c'}}
\label{triplephi}
\ee
where 
$f^{\textsf{abc}}$ and $\tilde f^{\textsf{a'b'c'}}$
are the structure constants of $U(N)$ and $U(\tilde{N})$.
We will refer to this as the {\it double-color theory}.
The partial amplitudes of this theory were shown 
in refs.~\cite{Cachazo:2013iea,Dolan:2013isa} to be equivalent to the 
propagator matrix (\ref{propagator}).
\para

In this section, we generalize the double-color theory to
include also scalar particles
$\psi^{\textsf{i}\,\textsf{i'}} $ 
in the $(fund,fund)$ representation of the group
$U(N) \times U(\tilde{N})$, 
with cubic couplings
\be
(T^{\textsf{a}})^{\textsf{i}}_{~\textsf{j}} 
(\tilde T^{\textsf{a'}})^{\textsf{i'}}_{~\textsf{j'}} 
\overline\psi_{\textsf{i}\,\textsf{i'}} \phi_{\textsf{a}\textsf{a'}}  \psi^{\textsf{j}\,\textsf{j'}}
\label{psiphipsi}
\ee
as well as a mass term
\be
m_\psi^2 
~\overline\psi_{\textsf{i}\,\textsf{i'}} 
\psi^{\textsf{i}\,\textsf{i'}} \,.
\ee
The tree-level amplitude for $(n-2)$ $\phi$ fields and two $\psi$ fields is
given by the sum over all cubic diagrams 
\be
\cAsc (\psiphirange) ~=~ \sum_i {c'_i ~ {\tilde c}'_i \over d'_i }
\label{ccd}
\ee
where $c'_i$, ${\tilde c}'_i$ are the color factors
constructed from the cubic vertices 
(\ref{triplephi}) and (\ref{psiphipsi}),
as discussed in sec.~\ref{sec:propagator},
and $d'_i$ is the product of massless $\phi$ and massive $\psi$ propagators.
\Eqn{ccd} can be rewritten
(again following the discussion in sec.~\ref{sec:propagator}) as
\be
\cAsc (\psiphirange) =  \sum_{\gamma, \, \delta \,  \in \, S_{n-2}} 
{\bf t}_{1 \gamma n } ~m'( 1 \gamma n| 1 \delta n) ~{\bf \tilde t}_{1 \delta n }  
\label{tmt}
\ee
where 
$m'( 1 \gamma n| 1 \delta n)$
is defined in \eqn{primepropagator}.
In the context of the double-color theory, 
the $m'( 1 \gamma n| 1 \delta n) $
play the role of double-partial amplitudes.
\para

We propose that the generalization of the CHY 
formulation to the double-color amplitude
with $n-2$ massless $\phi$ and two massive $\psi$ fields is\footnote{Our
overall sign convention differs from ref.~\cite{Cachazo:2013iea} 
in order that the double-partial amplitudes of the theory
will be precisely equal to the propagator matrix (\ref{primepropagator}).} 
\be
\cAsc (\psiphirange)
~=~
(-1)^{n-1}  ~\int \frac{d\,^n\sigma}{\textrm{vol}\,\SL2C}
{{\prod_a }'
~ \delta \left(\sum_{b\neq a} \frac{k_a \cdot k_b +\Delta_{ab} }{\sigma_a-\sigma_b} \right)
}
C(\sigma) \tilde C(\sigma)
\label{intCC}
\ee
where 
$\Delta_{ab} = m_\psi^2 (\delta_{a,1} \delta_{b,n} + \delta_{a,n} \delta_{b,1} )$
and
\be
C(\sigma) = 
\sum_{\gamma \in S_{n-2} }
\frac{  {\bf t}_{1 \gamma n} }
{\sigma_{1,\gamma(2)}\cdots\sigma_{\gamma(n-1),n}\sigma_{n,1}},
\qquad
\tilde C(\sigma) = 
\sum_{\gamma \in S_{n-2} }
\frac{  {\bf \tilde t}_{1 \gamma n} }
{\sigma_{1,\gamma(2)}\cdots\sigma_{\gamma(n-1),n}\sigma_{n,1}}
\label{defC}
\ee
with $\sigma_{ab} \equiv \sigma_a - \sigma_b$.
\Eqn{intCC} differs in two respects
from the corresponding expression in ref.~\cite{Cachazo:2013iea}
for the pure $\phi$ amplitudes:
(a) ${\bf c}_{1 \gamma n}$  is replaced by  
    ${\bf t}_{1 \gamma n}$ in the definition of $C(\sigma)$, and 
(b) the arguments of the delta functions
are the massive rather than the massless scattering equations.
\para

As explained in ref.~\cite{Cachazo:2013hca},
the delta functions completely localize the integral (\ref{intCC}).
Because of the linear dependence among the $n$ scattering 
equations (\ref{newscatt})
the delta functions for three of them ($a=i$, $j$, and $k$) may be omitted.
The expression appearing in \eqn{intCC} 
\be
\prod_a {}'
~\delta\left(\sum_{b\neq a} \frac{k_a \cdot k_b + \Delta_{ab} }{\sigma_a-\sigma_b}\right) 
\equiv \sigma_{ij}\sigma_{jk}\sigma_{ki}\prod_{a\neq i,j,k}
\delta\left(\sum_{b\neq a} \frac{k_a \cdot k_b  + \Delta_{ab}}{\sigma_a-\sigma_b}\right)
\ee
is independent of the choice of $i$, $j$, and  $k$.
Furthermore, because the integrand in \eqn{intCC} is $\SL2C$-invariant,
three of the $\sigma_a$
(arbitrarily chosen as $a=p$, $q$, and $r$)
can be fixed.
Including the Faddeev-Popov Jacobian that results from this, 
the integral (\ref{intCC}) evaluates to 
\be
\cAsc (\psiphirange)
= (-1)^{n-1} \sum_{\{\sigma\} \in {\rm solutions}}
\frac{ C(\sigma) \tilde C(\sigma) }{  \det'\Phi (\sigma)  }
\label{sumCC}
\ee
where the sum is over the $(n-3)!$ solutions of the scattering equations (\ref{newscatt}) 
and 
\be
{\det}'\Phi \equiv  \frac{|\Phi|^{ijk}_{pqr}}{(\sigma_{pq}\sigma_{qr}\sigma_{rp})(\sigma_{ij}\sigma_{jk}\sigma_{ki})}.
\label{detPhi}
\ee
Here $\Phi$ is a $n \times n$ matrix with entries 
\be
\Phi_{ab} = 
\frac{2(k_a \cdot k_b  + \Delta_{ab})}{(\sigma_a-\sigma_b)^2}, \quad  a\neq b;
\qquad\qquad 
\Phi_{aa} = 
 -\sum_{c\neq a}\frac{2(k_a \cdot k_c  + \Delta_{ac})}{(\sigma_a-\sigma_c)^2} 
\,.
\ee
This matrix has rank $(n-3)$
since $\sum_{a=1}^n \Phi_{ab} 
= \sum_{a=1}^n \sigma_a \Phi_{ab} 
= \sum_{a=1}^n \sigma_a^2 \Phi_{ab} = 0$
when the scattering equations are satisfied.
$\Phi^{ijk}_{pqr}$ is the nonsingular matrix obtained by removing rows 
$i$, $j$, and $k$, and columns $p$, $q$, and $r$,
and  $|\Phi|^{ijk}_{pqr}$ is its signed determinant.
Then $ \det'\Phi$ is independent of the choices of removed rows and columns.
\para

\Eqn{sumCC} implies that the massive double-partial amplitudes 
defined by \eqn{tmt} are given by 
\be
m'( 1 \gamma n | 1 \delta n) 
= \sum_{\{\sigma\} \in {\rm solutions}}\frac{(-1)^{n-1} }{\det'\Phi}
\frac{1}
{
(\sigma_{1,\gamma(2)}\cdots\sigma_{\gamma(n-1),n}\sigma_{n,1})
(\sigma_{1, \delta(2)}\cdots\sigma_{ \delta(n-1),n}\sigma_{n,1})
}
\label{doublepartial}
\ee
Each term in this sum is invariant under $\SL2C$,
so we may use an $\SL2C$ transformation to set $\sigma_1=0$, $\sigma_2=1$,
and $\sigma_n=\infty$ when evaluating \eqn{doublepartial}.
To give a concrete example, for $n=4$ we obtain
\ba
\begin{pmatrix}
m'( 1234 | 1234 ) & m'( 1234 | 1324 )  \\
m'( 1324 | 1234 ) & m'( 1324 | 1324 ) 
\end{pmatrix}
&=&
\begin{pmatrix}
{1 \over 2 k_2 \cdot k_3}
+{1 \over 2 k_3 \cdot k_4}
& 
-{1 \over 2 k_2 \cdot k_3}
 \\
-{1 \over 2 k_2 \cdot k_3}
& 
{1 \over 2 k_2 \cdot k_3}
+{1 \over 2 k_2 \cdot k_4}
 \\
\end{pmatrix}
\label{fourpointdoublepartial}
\ea
where in this case there is a single solution to the scattering equations
$\sigma_3 = -k_2 \cdot k_4/k_3 \cdot k_4$
on which $\det' \Phi  \to 2 (k_3 \cdot k_4)^3/ \left[\sigma_4^4 (k_2\cdot k_3) (k_2 \cdot k_4)
\right]$.
\para

To show that \eqn{intCC} correctly calculates the 
tree-level amplitude for $n-2$ massless $\phi$ and two massive $\psi$ fields,
we must establish that \eqns{primepropagator}{doublepartial} yield equivalent results
for all the double-partial amplitudes.
First observe that, when expressed in terms of $k_a \cdot k_b$ with $2 \le a <  b \le n$,
\eqn{doublepartial} is independent of $m_\psi$, and furthermore has exactly 
the same form as the analogous quantity for the massless $\phi$ amplitude in ref.~\cite{Cachazo:2013iea}.
{}This is apparent from the example (\ref{fourpointdoublepartial}), 
and can be verified in general from the massive scattering equations
and the definition of $\Phi$.
Moreover, as was shown in sec.~\ref{sec:propagator}, 
when expressed in terms of the same variables,
the massive propagator matrix (\ref{primepropagator}) is identical 
to the propagator matrix (\ref{primepropagator}) for the $n$-gluon theory.    
The equivalence between the double-partial amplitudes
of the massless $\phi$ amplitude and the propagator matrix for the $n$-gluon theory
was previously established in ref.~\cite{Cachazo:2013iea} 
(see also ref.~\cite{Dolan:2013isa}).
Thus, \eqn{intCC} is validated. 
\para

Next we turn to gauge-theory amplitudes involving massive scalar fields
in the fundamental representation.

\section{Gauge theory amplitudes}
\setcounter{equation}{0}
\label{sec:gauge}

Cachazo, He, and Yuan have presented a formula for the
tree-level $n$-gluon amplitude in arbitrary spacetime dimension
in terms of a sum over solutions of the massless scattering equations
\cite{Cachazo:2013hca,Cachazo:2013iea},
which was subsequently proved in ref.~\cite{Dolan:2013isa}.
In this section, we propose a CHY-type expression for the
gauge-theory amplitude for $(n-2)$ gluons and two massive 
scalars transforming in the fundamental representation, namely 
\ba
\cA (\psirange)
&=&
(-1)^{n-1}  ~\int \frac{d\,^n\sigma}{\textrm{vol}\,\SL2C}
{{\prod_a }'
~ \delta \left(\sum_{b\neq a} \frac{k_a \cdot k_b +\Delta_{ab} }{\sigma_{a,b}} \right)
}
C(\sigma) E(\sigma)
\nl
&=& (-1)^{n-1} \sum_{\{\sigma\} \in {\rm solutions}}
\frac{ C(\sigma) E(\sigma) }{  \det'\Phi (\sigma)  }
\label{sumCE}
\ea
which is obtained from the double-color amplitude (\ref{intCC}) 
presented in the previous section 
by simply replacing the factor $\tilde C(\sigma)$ with $E(\sigma)$.
{}Using \eqn{defC}, we find  that
the color-ordered amplitudes defined in \eqn{AtA} are given by
\be
A(1_\psi, \gamma(2),  \cdots, \gamma(n-1), n_\barpsi )  
 = 
\sum_{\{\sigma\} \in {\rm solutions}}\frac{(-1)^{n-1}  } {\det'\Phi}
\frac{E(\sigma)}
{ \sigma_{1,\gamma(2)}\cdots\sigma_{\gamma(n-1),n}\sigma_{n,1} } \,.
\label{colororderedCHY}
\ee
By substituting \eqn{doublepartial} into \eqn{AtA}
and comparing with \eqn{colororderedCHY}, we may deduce that
$E(\sigma)$ is related to the kinematic numerators via
\be
E(\sigma) = 
\sum_{\delta \in S_{n-2} }
\frac{  {\bf n'}_{1 \delta n} }
{\sigma_{1,\delta(2)}\cdots\sigma_{\delta(n-1),n}\sigma_{n,1}} \,.
\label{Enum}
\ee
The kinematic numerators depend on the momenta and polarizations of the external particles.
We now present an explicit expression for $E(\sigma)$ 
in terms of the pfaffian of an antisymmetric matrix $\Psi$.
\para

In refs.~\cite{Cachazo:2013hca,Cachazo:2013iea},
$\Psi$ is a $2n \times 2n$ matrix, 
in which the first $n$ entries correspond to the 
momenta of the gluons 
and the second $n$ entries correspond to 
their polarizations.
This matrix is singular, so to obtain a nonvanishing pfaffian 
it was necessary to remove two of the first $n$ rows and columns.
The choice of which rows/columns to remove was made arbitrarily 
and the result was shown to be independent of this choice.
In our case, there is the rather natural choice of removing 
the first and $n$th rows and columns, 
which are singled out as the momenta of the two massive scalars. 
Furthermore, we must also remove the $(n+1)$th and $(2n)$th rows
and columns, since the scalars have no polarizations. 
In our case, therefore, $\Psi$ is an antisymmetric
$(2n-4) \times (2n-4)$ matrix 
\be
\Psi = \left(
         \begin{array}{cc}
           A &  -C^{\rm T} \\
           C & B \\
         \end{array}
       \right)
\label{defPsi}
\ee
where $A$, $B$ and $C$ are the $(n-2)\times (n-2)$ submatrices\footnote{
We have removed the $\Delta_{ab}$'s that were present in the entries of $B$ and $C$ in 
{\tt v1} of this paper.  
Their presence is innocuous for the amplitudes considered in this paper
(in which only $\Delta_{1n}=\Delta_{n1}$ is nonvanishing), 
but the derivation in ref.~\cite{Naculich:2015} shows that
they are generally absent.
}
\ba
A_{ab} &=& \begin{cases} 
\displaystyle \frac{2(k_a\cdot k_b+\Delta_{ab}) }{\sigma_{a}-\sigma_{b}}, & a\neq b;\\[5mm]
\displaystyle \quad ~~ 0, & a=b;\end{cases} 
\quad\quad 
B_{ab} = \begin{cases} 
\displaystyle \frac{2 \epsilon_a\cdot \epsilon_b}{\sigma_{a}-\sigma_{b}}, & a\neq b;\\[5mm]
\displaystyle \quad ~~ 0, & a=b;\end{cases}
\ea
\ba
C_{ab} &=& \begin{cases} \displaystyle \frac{2 \epsilon_a\cdot k_b}{\sigma_{a}-\sigma_{b}}, &\quad a\neq b;\\[5mm]
\displaystyle -\sum_{c\neq a}\frac{2 \epsilon_a\cdot k_c}{\sigma_{a}-\sigma_{c}}, &\quad a=b;\end{cases}
\label{submatrices}
\ea
where the range of $a$ and $b$ is restricted to $2, \cdots, n-1$.
(The sum over $c$ in $C_{aa}$, however, runs from 1 to $n$, omitting $a$.)
Thus our $\Psi$ is simply a truncated version of the $\Psi$ 
defined in ref.~\cite{Cachazo:2013hca,Cachazo:2013iea}, modified to include masses.
\para

One may verify that the pfaffian of $\Psi$ is gauge-invariant 
(i.e., under $\epsilon_a \to \epsilon_a + k_a$)
when evaluated on a solution of the scattering equations.
We now define
\be
E(\sigma)  = \frac{1}{\sigma_{1n}^2} {\Pf} \Psi 
\ee
which implies that 
$E(\sigma) \to E(\sigma) \prod_{a=1}^{n} (C \sigma_a+ D)^2 $
under an $\SL2C$ transformation (\ref{sl2c}),
the correct behavior to ensure that each term in \eqn{colororderedCHY}
is $\SL2C$-invariant as well as gauge-invariant.
\para

We now evaluate \eqn{colororderedCHY}
for specific values of $n$. 
It is convenient to define the $\SL2C$-invariant expression
\be 
\hat E (\sigma) = \sigma_{12} \sigma_{23} \cdots \sigma_{n-1,n} \sigma_{n1} E(\sigma) \,.
\ee
For example, for $n=4$, we obtain\footnote{after using 
$\sum_{b\neq a} \epsilon_a \cdot k_b =  - \epsilon_a \cdot k_a =0$}
\be
\hat  E (\sigma)
=
-4 \left[ 
\epsilon_2 \cdot k_1 ~ \epsilon_3 \cdot k_4
- 
{\sigma_{12} \sigma_{34} \over \sigma_{13} \sigma_{24} }
\epsilon_2 \cdot k_4 ~ \epsilon_3 \cdot k_1 
+ 
{\sigma_{12} \sigma_{34} \over \sigma_{23} \sigma_{14} }
 k_2 \cdot k_3 ~\epsilon_2 \cdot \epsilon_3 
\right], \qquad n=4 
\label{Efour}
\ee
expressed in terms of $\SL2C$-invariant cross ratios.
Then, evaluating this on the single solution of the scattering equations
$\sigma_1=0$, $\sigma_2=1$, 
$\sigma_3 = -k_2 \cdot k_4/k_3 \cdot k_4$, and $\sigma_4 \to \infty$,
we obtain 
\ba
A(1_\psi, 2, 3, 4_\barpsi )  &=&
 {2  k_2 \cdot k_4 \over k_2 \cdot k_3 }
\left[ 
{  \epsilon_2 \cdot k_1 ~\epsilon_3 \cdot k_4 \over ~k_3 \cdot k_4 }
+ { \epsilon_2 \cdot k_4 ~ \epsilon_3 \cdot k_1 \over k_2 \cdot k_4 }
+ { \epsilon_2 \cdot \epsilon_3 }
\right]
\nl
&=& 
{4 (u- m_\psi^2) \over t} 
\left[
   {\epsilon_2 \cdot k_1 ~\epsilon_3 \cdot k_4 \over s-m_\psi^2  }~
+  {\epsilon_2 \cdot k_4 ~ \epsilon_3 \cdot k_1  \over u-m_\psi^2}
+ {1\over 2}~  \epsilon_2 \cdot \epsilon_3 
\right]
\ea
which is in agreement with a direct Feynman diagram evaluation.
\para

Various tree-level amplitudes for gluons and massive scalars have been 
calculated in four dimensions 
using recursive techniques ~\cite{Bern:1996ja,Badger:2005zh,Forde:2005ue}.
We have numerically evaluated \eqn{colororderedCHY} 
for $n=5$ and $n=6$
and have obtained agreement with these  results (up to overall normalization).
Specifically, we have verified that for $n=5$ 
with various helicity configurations
\eqn{colororderedCHY} yields\footnote{Note the change in 
sign in the second term of $A(1_\psi, 2^+, 3^+, 4^-, 5_\barpsi )$
relative to ref.~\cite{Badger:2005zh}.
This correction was also noted in ref.~\cite{Forde:2005ue}.}
\ba
A(1_\psi, 2^+, 3^+, 4^+, 5_\barpsi )  &=& 
2 \sqrt{2} 
\left[ m_\psi^2 \left( [42] \langle 2| k_1 | 2] + [43] \langle 3| k_1 | 2] \right) 
\over ( 2 k_1 \cdot k_2 ) \langle 23 \rangle  \langle 34 \rangle (2 k_4 \cdot k_5) \right],
\label{spinorfiveA}
\\ [3mm]
A(1_\psi, 2^+, 3^+, 4^-, 5_\barpsi )  &=& 
2 \sqrt{2}  
\bigg[
- { \left( \langle 4 | k_5 | 2]  \langle 2| k_1 | 2] + \langle 4 | k_5 | 3]  \langle 3| k_1 | 2] \right)^2
	\over ( 2 k_1 \cdot k_2 ) \langle 23 \rangle  \langle 34 \rangle (2 k_4 \cdot k_5) 
	\left( [42] \langle 2| k_1 | 2] + [43] \langle 3| k_1 | 2] \right) }
\nl
&&~~~~~~~~ + 
{ m_\psi^2 [23]^3 \over
\left( k_2 + k_3 + k_4 \right)^2 [34] \left( [42] \langle 2| k_1 | 2] + [43] \langle 3| k_1 | 2] \right) }
\bigg],
\label{spinorfiveB}
\\ [3mm]
A(1_\psi, 2^+, 3^-, 4^+, 5_\barpsi )  &=& 
2 \sqrt{2}  
\bigg[
- { \langle 3 | k_1 | 2]^2   \langle 3 | k_5 | 4]^2 \over
	( 2 k_1 \cdot k_2 ) \langle 23 \rangle  \langle 34 \rangle (2 k_4 \cdot k_5) 
	\left( [42] \langle 2| k_1 | 2] + [43] \langle 3| k_1 | 2] \right) }
\nl
&&~~~~~~~~ + 
{ m_\psi^2 [24]^4 \over
\left( k_2 + k_3 + k_4 \right)^2 [23] [34] \left( [42] \langle 2| k_1 | 2] + [43] \langle 3| k_1 | 2] \right) }
\bigg].
\nl
\label{spinorfiveC}
\ea
For $n=6$, we have verified that \eqn{colororderedCHY} gives
\ba
&& A(1_\psi, 2^+, 3^+, 4^+, 5^+, 6_\barpsi ) 
\nl
&&
~~~~~~=4
\left[ 
- m_\psi^2 
 [5 | k_6 (4+5) (2+3) k_1 | 2]
\over 
( 2 k_1 \cdot k_2 ) 
( 2 k_1 \cdot k_2  + 2 k_1 \cdot k_3 + 2 k_2 \cdot k_3) 
( 2 k_5 \cdot k_6 )
\langle 23 \rangle  
\langle 34 \rangle 
\langle 45 \rangle 
\right],
\label{spinorsix}
\ea
and have also checked agreement with 
$ A(1_\psi, 2^+, 3^+, 4^+, 5^-, 6_\barpsi ) $
in ref.~\cite{Badger:2005zh}.
\para

While generically all $(n-3)!$ solutions of the scattering equation 
contribute to the amplitude (\ref{colororderedCHY}),
we have observed that,
in the massless limit $m_\psi \to 0$ in four dimensions,
only a subset of solutions contributes
to any given amplitude.
For amplitudes in which all gluons have the same helicity,
the amplitude vanishes as $m_\psi \to 0$, 
and in fact the contribution from each solution
of the scattering equations individually vanishes.
For amplitudes in which all gluons but one 
have the same helicity (MHV amplitudes),
only one of the solutions of the scattering equations 
contributes to the amplitude.
More precisely,
for the mostly-plus helicity amplitude,
only the solution (cf. ref.~\cite{Weinzierl:2014vwa})
\be
\sigma_i = 
{\langle i 1 \rangle \langle  2 n \rangle \over
 \langle i n \rangle \langle  2 1 \rangle }, \qquad i = 1, \cdots, n
\ee
contributes,
while for the mostly-minus helicity amplitude, 
only the solution 
\be
\sigma_i = 
{[ i 1 ] [  2 n ] \over
 [ i n ] [  2 1 ] }, \qquad i = 1, \cdots, n
\ee
contributes.
For $n=6$, neither of these solutions contributes to the amplitudes
with two positive and two negative helicities, 
but all of the other four solutions do.
We expect these patterns to continue to higher values of $n$.

\subsection{BCJ relations from scattering equations}

The BCJ relations follow from color-kinematic duality and the 
reduced rank of the propagator matrix.
As shown in ref.~\cite{Cachazo:2013iea},
a gauge-theory amplitude that is written in the CHY form automatically 
satisfies these properties
and consequently, the color-ordered amplitudes must satisfy BCJ relations.
Here we will explicitly demonstrate that the CHY-type expression (\ref{colororderedCHY})
for the color-ordered amplitudes for $(n-2)$ gluons and two massive scalars
obeys the fundamental BCJ relation for massive amplitudes (\ref{intromassivebcj}).
Although this argument has essentially been given already 
in refs.~\cite{Cachazo:2012uq,Cachazo:2013iaa},
we repeat it here to be self-contained.
\para

The fundamental BCJ relation (\ref{massivebcj})
can be recast as 
\be
0 = \sum_{b=3}^n   k_2 \cdot k_b 
\sum_{a=3}^b A(1_\psi, 3, \cdots, a-1, 2, a, \cdots, n_\barpsi) \,.
\ee
We use \eqn{colororderedCHY} to compute
\ba
&& 
\sum_{a=3}^b A(1_\psi, 3, \cdots, a-1, 2, a, \cdots, n_\barpsi)
\nl
&=&
 \sum_{\{\sigma\} \in {\rm solutions}}\frac{ (-1)^{n-1} } {\det'\Phi}
\frac{E(\sigma)}
{  \sigma_{13} \sigma_{34} \cdots \sigma_{n-1,n} \sigma_{n1}  }
\left[  { \sigma_{13} \over \sigma_{12} \sigma_{23} } + 
\sum_{a=4}^b { \sigma_{a-1,a} \over \sigma_{a-1,2} \sigma_{2a} } \right]
\nl
&=& \sum_{\{\sigma\} \in {\rm solutions}}\frac{ (-1)^{n-1} } {\det'\Phi}
\frac{E(\sigma)}
{  \sigma_{13} \sigma_{34} \cdots \sigma_{n-1,n} \sigma_{n1}  }
{ \sigma_{1b} \over \sigma_{12} \sigma_{2b}  } \,.
\ea
Now consider
\ba
\sum_{b = 3}^n { k_2 \cdot k_b} 
{ \sigma_{1b} \over \sigma_{12} \sigma_{2b}  }
&= & \sum_{b\neq 2}  { k_2 \cdot k_b} 
{ \sigma_{1b} \over \sigma_{12} \sigma_{2b}  }
~=~
\sum_{b \neq 2}  { k_2 \cdot k_b} 
{ \sigma_{12} + \sigma_{2b} \over \sigma_{12} \sigma_{2b}  }
\nl
&=&
\sum_{b\neq 2} 
{ k_2 \cdot k_b 
\over \sigma_{2b}  }
+ {1 \over \sigma_{12}} \sum_{b\neq 2} 
{ k_2 \cdot k_b} 
~=~
\sum_{b\neq 2} 
{ k_2 \cdot k_b 
\over \sigma_{2b}  }
\ea
where the last equality follows by momentum conservation and $m_2^2=0$.
Then 
\ba
&&\sum_{b=3}^n   k_2 \cdot k_b 
\sum_{a=3}^b A(1_\psi, 3, \cdots, a-1, 2, a, \cdots, n_\barpsi)
\nl
&=& \sum_{\{\sigma\} \in {\rm solutions}}\frac{(-1)^{n-1}  } {\det'\Phi}
\frac{E(\sigma)}
{ 
 \sigma_{13} \sigma_{34} \cdots \sigma_{n-1,n} \sigma_{n1}  }
\left( \sum_{b\neq 2} 
{ k_2 \cdot k_b 
\over \sigma_{2b}  } \right)
= 0
\ea
because the term in parentheses vanishes on any solution of the scattering equations
(\ref{middle}).

\section{Gravitational amplitudes} 
\setcounter{equation}{0}
\label{sec:gravity}

Cachazo, He, and Yuan have presented a formula for the
tree-level $n$-graviton amplitude in arbitrary spacetime dimension
in terms of a sum over solutions of the massless scattering equations
\cite{Cachazo:2013hca,Cachazo:2013iea}.
In this section, we propose an analogous expression for the
amplitude for $(n-2)$ gravitons and two massive 
scalars in terms of solutions of the massive scattering equations, namely 
\ba
\cAgv (\psirange)
&=&
(-1)^{n-1}  ~\int \frac{d\,^n\sigma}{\textrm{vol}\,\SL2C}
{{\prod_a }'
~ \delta \left(\sum_{b\neq a} \frac{k_a \cdot k_b +\Delta_{ab} }{\sigma_{a,b}} \right)
}
E(\sigma) \tilde E(\sigma)
\nl 
&=& (-1)^{n-1} \sum_{\{\sigma\} \in {\rm solutions}}
\frac{ E(\sigma) \tilde E(\sigma) }{  \det'\Phi (\sigma)  }
\label{sumEE}
\ea
which is obtained from the gauge-theory amplitude (\ref{sumCE}) 
presented in the previous section 
by replacing the factor $C(\sigma)$ with $\tilde E(\sigma)$.
All the ingredients in this equation have already been defined in previous
sections.
\para

For $n=4$, the two-graviton two-scalar scattering amplitude 
was computed long ago \cite{Gross:1968in};
the calculation is rather tedious.
In contrast, it is trivial to evaluate \eqn{sumEE} on the single solution 
of the scattering equations
$\sigma_1=0$, $\sigma_2=1$, 
$\sigma_3 = -k_2 \cdot k_4/k_3 \cdot k_4$, and $\sigma_4 \to \infty$
and use \eqn{Efour} to obtain
\ba
\cAgv (1_\psi, 2, 3, 4_{\barpsi})
&=&
~-~ { 8 k_2 \cdot k_4  ~k_3 \cdot k_4 \over k_2 \cdot k_3} 
\left[ 
{  \epsilon_2 \cdot k_1 ~\epsilon_3 \cdot k_4 \over ~k_3 \cdot k_4 }
+ { \epsilon_2 \cdot k_4 ~ \epsilon_3 \cdot k_1 \over k_2 \cdot k_4 }
+ { \epsilon_2 \cdot \epsilon_3 }
\right]
\nl
&& ~~~~~~~~~~~~~\times
\left[ 
{  \tep_2 \cdot k_1 ~\tep_3 \cdot k_4 \over ~k_3 \cdot k_4 }
+ { \tep_2 \cdot k_4 ~ \tep_3 \cdot k_1 \over k_2 \cdot k_4 }
+ { \tep_2 \cdot \tep_3 }
\right] \,.
\ea
This has the nice feature of yielding the known factorized form for the 
amplitude \cite{Choi:1994ax,Holstein:2006bh}.
\para

We hope that the expression (\ref{sumEE}) may play a useful role in the
current lively discussion of soft graviton 
theorems \cite{Cachazo:2014fwa,Schwab:2014xua,Afkhami-Jeddi:2014fia,Zlotnikov:2014sva,Kalousios:2014uva}.
\para

\section{Conclusions}
\setcounter{equation}{0}
\label{sec:concl}

In this paper, we have examined color-kinematic duality for gauge theories 
with massive particles in representations other than the adjoint.
We derived the form of the BCJ relations for tree-level amplitudes 
with $n-2$ gluons and a pair of massive particles
that are implied by color-kinematic duality.
\para

We have also generalized the scattering equations to include both massless
and massive particles,
and have proposed  CHY-type expressions 
for tree-level amplitudes in three interrelated theories
in terms of the solutions to these equations: 
\ba
\cAsc (\psiphirange)
&=&
 (-1)^{n-1} \sum_{\{\sigma\} \in {\rm solutions}}
\frac{ C(\sigma) \tilde C(\sigma) }{  \det'\Phi (\sigma)  }\,,
\nl
\cAga (\psirange)
&=& (-1)^{n-1} \sum_{\{\sigma\} \in {\rm solutions}}
\frac{ C(\sigma) E(\sigma) }{  \det'\Phi (\sigma)  }\,,
\label{CHY}
\\
\cAgv (\psirange)
&=& (-1)^{n-1} \sum_{\{\sigma\} \in {\rm solutions}}
\frac{ E(\sigma) \tilde E(\sigma) }{  \det'\Phi (\sigma)  } \, .
\nonumber
\ea
Particles 2 through $n-1$ are massless scalars, gluons, and gravitons
respectively, 
and particles 1 and $n$ are in each case massive scalars.
The summand of each of these expressions is a product of factors:
$C(\sigma)$ represents the color factor (\ref{defC}) and 
$E(\sigma)$ represents a polarization-dependent kinematic factor (\ref{Enum}), 
which can be written in terms of the pfaffian of a matrix (\ref{defPsi}).
Color-kinematic duality of the gauge theory 
and the double-copy prescription for gravity
are completely manifest in these expressions.
\para

When the scattering equations have only one solution
(e.g. for four-point amplitudes),
\eqn{CHY} implies that the amplitude itself 
can be expressed as a product of factors.
This neatly explains the  factorization of various four-point amplitudes 
observed in gauge theory \cite{Zhu:1980sz,Goebel:1980es}
and gravity \cite{Choi:1994ax,Holstein:2006bh}.
Conversely, the fact that gauge-theory and gravity 
four-point amplitudes involving massive fermions 
also factorize 
strongly hints that expressions analogous to \eqn{CHY}
should also exist for fermions.
\para

In cases where only one solution of the scattering
equations contributes to the amplitude,
e.g., MHV amplitudes in four dimensions with $m_\psi=0$,
\eqn{CHY} again implies that the amplitude should factorize.
This observation could lead to simpler expressions for this class of 
amplitudes.
\para

Finally, an obvious and very important direction 
for future research 
is the generalization of the scattering equation approach
to loop-level amplitudes.

\section*{Acknowledgments}
I am grateful to Freddy Cachazo, Louise Dolan, Henriette Elvang, 
Michael Kiermaier, and Ellis Yuan 
for useful conversations.
This research was supported in part by the NSF under grant no. PHY10-67961

\vfil\break

\end{document}